\documentclass[%
 reprint,
runinaddress,
 amsmath,amssymb,
 aps,
prresearch,
]{revtex4-2}

\usepackage{graphicx}% Include figure files
\usepackage{dcolumn}% Align table columns on decimal point
\usepackage{bm}% bold math
\usepackage{epstopdf, epsfig}
\usepackage{tikz}

\usepackage{verbatim}
\usepackage{amsmath}%
\usepackage{longtable}%
\usepackage{textcomp}
\usepackage{color}
\usepackage[normalem]{ulem}

\begin{document}

\title{Transition to dilatation-dominated isothermal compressible turbulence}

\author{Shadab Alam$^{1}$}
\author{Christoph Federrath$^{2}$}
\author{J\"org Schumacher$^{1}$}
\email{joerg.schumacher@tu-ilmenau.de}
\affiliation{$^{1}$Institut f\"ur Thermo- und Fluiddynamik, Technische Universit\"at Ilmenau, P.O.Box 100565, D-98684 Ilmenau, Germany\\
$^{2}$Research School of Astronomy and Astrophysics, Australian National University, Canberra, ACT 2611, Australia}

\date{\today}

\begin{abstract}
The kinetic energy dissipation rate is of central importance for small-scale statistics in turbulent flows. Here, we report a transition to the dilatation-dominated regime of three-dimensional isothermal  fully compressible, homogeneous, isotropic turbulence by moments of energy dissipation and its components up to order~4 for root-mean-square (rms) Mach numbers $0.1\le M_{\rm rms}\le 10$ and for Reynolds numbers $100\le Re\le 2400$. Our high-resolution numerical simulations show a crossover from incompressible to $M_{\rm rms}$--independent, Burgers turbulence-like scaling of energy dissipation rate moments with respect to Reynolds number $Re$. This confirms the statistical dominance of shocks for rms Mach numbers $M_{\rm rms}\gtrsim 1$.    
\end{abstract}

\maketitle

\section{Introduction}
Compressible turbulence introduces additional layers of complexity through mechanisms  absent in its incompressible counterpart, including shocklets and shocks, dilatational strain and  thermodynamic interactions~\citep{DJ2013, FederrathOffner2025}. These mechanisms shape the flow dynamics in many systems ranging from the interstellar medium (ISM) to engineering applications, such as gas turbines and supersonic propulsion. This is done by modifying the energy cascade~\citep{aluie2011comp,Galtier_2011,WYS+2013,Federrath2013,Kritsuk_2013,ED2018} and thus influencing small-scale statistics and dissipation of kinetic energy~\citep{PPK2009, JDSJFM2021}. The root-mean-square (rms) Mach number and large-scale Reynolds number, which are given by
\begin{equation}
M_{\rm rms} = \frac{u_{\rm rms}}{c_s}\quad\mbox{and}\quad Re = \frac{u_{\rm rms} L_f}{\nu}\,,
\end{equation}
characterize, respectively, the level of compressibility and the vigor of turbulence. Here, $\nu$ is the kinematic viscosity, $L_f$ the forcing length scale, $c_s$ the speed of sound, $u_{\rm rms} = \sqrt{\langle {\bm u}^2 \rangle}$ is rms of velocity field ${\bm u}$, and $\langle \cdot \rangle$ denotes the average over volume $V$ and time $t$. The flow is homogeneous and locally isotropic with $\langle {\bm u} \rangle\approx 0$. 

A defining characteristic of fully turbulent flows is spatial intermittency--the emergence of strongly localized bursts, steep fronts and intense vortices at the small-scale end of the inertial cascade close to the Kolmogorov dissipation length $\eta$ \cite{She1991,Schumacher2010,Iyer2018,Buaria:NJP2019,Buaria2024,Sreenivasan:ARCMP2025}. They give rise to strongly non-Gaussian statistics of turbulence fields and their derivatives, in particular the kinetic energy dissipation rate $\epsilon$, which is given by
\begin{equation}
	\epsilon({\bm x}, t) = 2 \mu {\bm S}: {\bm S} - \frac{2 \mu}{3} ({\bm \nabla} \cdot {\bm u})^2\,,
	\label{eq:eps_total_1}
\end{equation}
with the dynamic viscosity $\mu$ and the rate-of-strain tensor ${\bm S} =[{ \bm \nabla} {\bm u} + ( {\bm \nabla} {\bm u})^{T}]/2$, which probes the strength of local shear. In the inertial range for scales $\ell>\eta$ and at $M_{\rm rms} \le 1$, the structure functions of the velocity field were found to follow scaling laws with power law exponents $\zeta_n$ close to incompressible predictions \citep{Wang:PRL2012, Wang:2017_intermittency}. In Lagrangian studies of compressible turbulence, the situation is slightly different. For $M_{\rm rms} \lesssim 0.5$, the statistics remain essentially incompressible~\citep{Benzi_2010, Yang_2016}; however, deviations become evident at higher $M_{\rm rms}$~\citep{Konstandin2012,Yang_2016} as shocklets begin to influence the dynamics. In ref.~\cite{PJNB2004}, these scaling exponents $\zeta_n$ were shown to fit to a She-L\'ev\^eque-type formula for the compressible case, where a single parameter--the Hausdorff dimension (or box counting dimension) $D$ of the most intense dissipative structures--varies with $M_{\rm rms}$. Supersonic isothermal simulations at $M_{\rm rms} = 6$ and 17 \cite{KNPW2007,Federrath2013} exhibited for example kinetic energy spectra close to the shock-dominated Burgers turbulence-type $k^{-2}$ scaling, together with larger values of the low-order structure-function exponents~\citep{KNPW2007} compared to incompressible turbulence. The deviations increase when the volume forcing of the compressible flow injects additional velocity field divergence into the flow \cite{SFK2008,Konstandin2012,Federrath2013}, i.e., when it becomes increasingly dilatational. These results clearly reflect the growing impact of shocks on the turbulence statistics. As we can see, most analysis of compressible turbulence focused on the statistics in the inertial cascade range, such as scaling laws for energy spectra and structure functions at sufficiently high Reynolds numbers.
%---------------------------------------------
\begin{figure*}[htbp!]
\centering
\includegraphics{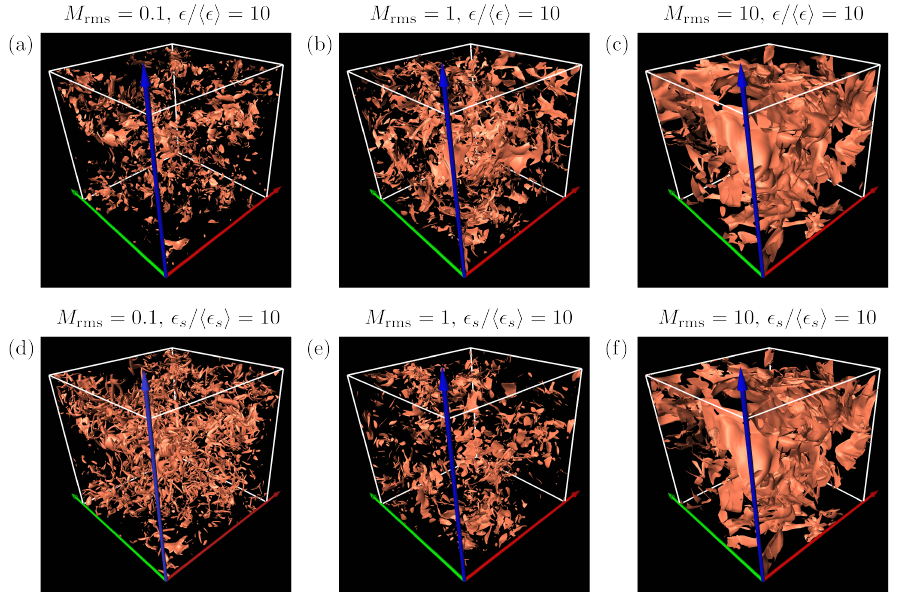}
\caption{Structure of the kinetic energy dissipation rate field. Isosurface snapshots of total energy dissipation $\epsilon({\bm x}, t_0)$ at the level $\epsilon = 10 \langle \epsilon \rangle$ (top row) and its solenoidal component $\epsilon_s({\bm x}, t_0)$ at $\epsilon_s = 10 \langle \epsilon_s \rangle$ (bottom row) for rms Mach numbers $M_{\rm rms} = 0.1$ (a, d), $M_{\rm rms} = 1$ (b, e) and $M_{\rm rms} = 10$ (c, f). All data are obtained at the highest Reynolds number $Re \approx 2400$; isosurfaces are displayed for the decadic logarithm of the fields.}
\label{fig:isosurface}
\end{figure*}
%---------------------------------------------

Here, we will turn the attention from the velocity to the velocity derivative statistics at somewhat smaller Reynolds numbers. In these cases, a developed inertial cascade range with a power law scaling of structure functions is absent, but we will show that the intermittent statistical properties of derivatives have already developed, e.g., for the kinetic energy dissipation rate and, in compressible flow case, its solenoidal and dilatational components. It is known from incompressible flows that derivative moments transition from Gaussian to intermittent scaling at $Re\ \sim 10^2$ \cite{yakhot2006,schumacher2007,SSY07,SSK+2014,Donzis2017,JS2018,Sreenivasan2021,Gotoh2022}. The same holds for Burgers turbulence \cite{Friedrich2018}. Connections between turbulent statistics at different scales are described by fusion rules. They either connect velocity structure functions at different distances $R>r$ in multiscale correlators in the inertial range \cite{BBT1998} or velocity structure functions to the Reynolds number scaling of dissipation rate moments, which in case of the energy dissipation moments for incompressible turbulence leads to~\cite{yakhot2006} 
\begin{equation*}
    M_{n} = \frac{ \langle \epsilon^{n} \rangle } {\langle \epsilon \rangle^{n}} \propto Re^{\beta_{n}}\;\;\mbox{with}\;\;\beta_n = n+\frac{\zeta_{4n}}{\zeta_{4n} - \zeta_{4n+1}-1}.    
    \label{eq:moments}
\end{equation*}
The scaling exponents $\beta_{n}$ depend nonlinearly on the order $n$ (via the structure function exponents $\zeta_n$), thus reflecting the intermittent character of the fields by an anomalous moment scaling with respect to $Re$ \cite{yakhot2006}. 

Here, we conduct a series of direct numerical simulations (DNS) of compressible turbulence at very high resolution. We monitor the Reynolds number scaling of $M_n(Re)$ for different degrees of compressibility, quantified by the rms Mach number $M_{\rm rms}$. Our main finding is a detection of a transition from a solenoidal statistics of moments $M_n$ for $M_{\rm rms}\lesssim 1$ to a close-to-Burgers-turbulence-type behavior for $M_{\rm rms} > 1$, where the energy dissipation is dominated by its dilatational dynamics.     

\section{DIRECT NUMERICAL SIMULATIONS} 
The high-resolution DNS of three-dimen\-sional homogeneous, isotropic, compressible turbulence solve the following equations,
\begin{align}
	\frac{\partial \rho}{\partial t}  +  {\bm \nabla}  \cdot \left( {\rho \bm u} \right) = 0
	\label{eq:mass}\,,\\
	\frac{\partial (\rho \bm u)}{\partial t} + {\bm \nabla}  \cdot  \left( \rho {\bm u} \otimes {\bm u} \right) &= - {\bm \nabla} p  + {\bm \nabla} \cdot {\bm \sigma} + \rho {\bm f}
	\label{eq:mom}\,,
\end{align}
together with the isothermal equation of state, $p = \rho c_s^2$.
Here, $\rho$ is the mass density, $p$ the pressure, ${\bm \sigma}$ the viscous stress tensor, $\bm f$ an external forcing, and $\otimes$ denotes the dyadic product. The isothermal closure corresponds to a barotropic model, $p= K \rho$, which is not thermodynamically faithful for isolated or weakly cooled molecular fluids, but can approximate rapidly cooled systems where heat generated by viscous dissipation is removed to keep the temperature nearly constant~\cite{ED2018}. Such conditions occur when the cooling time is much shorter than fluid time scales, such as in the dense, star-forming phase of the interstellar medium. The viscous stress tensor is given by
\begin{align}
{\bm \sigma}
=
\mu
\left[
{\bm \nabla} {\bm u}
+
({\bm \nabla} {\bm u})^{T}
\right]
+
\left(
\zeta - \frac{2\mu}{3}
\right)
({\bm \nabla} \cdot {\bm u})
{\bm I},
\end{align}
where ${\bm I}$ denotes the identity tensor. We set the bulk viscosity $\zeta = 0$, corresponding to a Newtonian fluid under the Stokes hypothesis. The flow is driven by a stochastic forcing ${\bm f}$ implemented in Fourier space using an Ornstein--Uhlenbeck (OU) process \citep{Federrath:2010,FederrathEtAl2022ascl}. The forcing is {\em purely solenoidal}, applied at the smallest wavenumbers, $1 < k\,L/2\pi < 3$, with a parabolic spectral distribution peaking at $k\,L/2\pi = 2$, where $L$ is the domain size. This corresponds to a characteristic forcing length scale $L_f \approx L/2$. The equations are solved in a triply periodic cube of volume $V=L^3$ using a modified version of the \texttt{FLASH} code~\citep{Fryxell:2000,Dubey:2008,FederrathOffner2025}. In each simulation run, the kinematic viscosity $\nu = \mu / \rho$ is kept constant, while the dynamic viscosity $\mu = \rho \nu$ varies with the local density. The sound speed is fixed at $c_s = 1$, while the mean kinetic energy injection rate $\langle \epsilon_{\rm in} \rangle = \langle \rho {\bm f} \cdot {\bm u} \rangle$ is varied to attain the target $M_{\rm rms}$. For each fixed $M_{\rm rms}$, different Reynolds number $Re$ are obtained by adjusting $\nu$ between runs.

Our DNS cover two orders of magnitude of rms Mach numbers, $0.1 \le M_{\rm rms} \le 10$, thus spanning subsonic and supersonic regimes. DNS are performed at very high spatial resolutions with $k_{\rm max}\eta > 10$, where $k_{\rm max} = \pi N/L$ and the Kolmogorov length scale $\eta = (\langle \rho \rangle \nu^3 / \langle \epsilon \rangle)^{1/4}$. We conducted 56~simulations on uniform meshes with up to $N^3=2048^3$ grid points. The attainable Reynolds numbers are $100 \lesssim Re \lesssim 2400$ \citep{ShivakumarFederrath2025}, with corresponding Taylor microscale Reynolds numbers $15 \lesssim Re_{\lambda} =\sqrt{(5 \langle \rho \rangle/(3 \nu \langle \epsilon \rangle)} u_{\rm rms}^2\lesssim 117$. All DNS thus start beyond the transition from Gaussian to non-Gaussian derivative statistics \cite{Donzis2017}. For each simulation, $N_s$ snapshots of the statistically stationary state are taken at intervals of one quarter of the large-scale eddy turnover time $L_f / u_{\rm rms}$ for statistical analysis. Details of the DNS parameters, characteristic statistical quantities, and numerical method are provided in Appendix~\ref{appen:numerical}. We also give detailed validations of our statistical results by comparison with other numerical methods for the lowest \cite{SSY07,SSK+2014,DJ2013,JD2016} and largest $M_{\rm rms}$ \cite{PPK2009} in Appendix~\ref{appen:validation}. There, we show that the present numerical scheme is not significantly affected by numerical artifacts for $M_{\rm rms}\ge 0.1$, see \cite{GV1999,GM2004,D2010}. 

Similar to the Helmholtz decomposition of the velocity field into solenoidal and dilatational parts, ${\bm u} = {\bm u}_s + {\bm u}_d$, with ${\bm \nabla}\cdot{\bm u}_s = 0$, and ${\bm \nabla}\times{\bm u}_d = 0$, we can decompose the kinetic energy dissipation rate field \eqref{eq:eps_total_1} into solenoidal, dilatational, and inhomogeneous components \cite{Huang:JFM1995,JDS2019,Alam_2025}. To derive this decomposition, we first write Eq.~\eqref{eq:eps_total_1} in index notation as
\begin{align}
\epsilon({\bm x},t)  &=  2\mu S_{ij}S_{ij}  - \frac{2\mu}{3} \left( \frac{
\partial u_k}{\partial x_k} \right)^2 \,.
\label{eq:viscous_dissi}
\end{align}
The first term on the right-hand side of (\ref{eq:viscous_dissi}) can be expanded as follows,
\begin{align}
2\mu S_{ij}S_{ij} = 2\mu \Omega_{ij}\Omega_{ij} + 2\mu \left[ \frac{\partial u_i}{\partial x_j}  \frac{\partial u_j}{\partial x_i} \right] \,,
\label{eq:sijsij}
\end{align}
where $\Omega_{ij} = (\partial u_{i}/\partial x_j-\partial u_{j}/\partial x_i)/2$ is the anti-symmetric vorticity tensor and $S_{ij} \Omega_{ij} = 0$. Using the product rule, we express the second term on the right hand side of (\ref{eq:sijsij}) as
\begin{align}
2\mu \frac{\partial u_i}{\partial x_j}\frac{\partial u_j}{\partial x_i}
&= 2\mu \left[ \frac{\partial}{\partial x_j}
\left( u_i \frac{\partial u_j}{\partial x_i} \right)
- u_i \frac{\partial}{\partial x_j}
\left( \frac{\partial u_j}{\partial x_i} \right) \right]
\nonumber\\
&= 2\mu \frac{\partial}{\partial x_j}
\left[ \frac{\partial}{\partial x_i}(u_i u_j)
- u_j \frac{\partial u_i}{\partial x_i} \right]
\nonumber\\
&\quad
- 2\mu \left[ \frac{\partial}{\partial x_i}
\left( u_i \frac{\partial u_j}{\partial x_j} \right)
- \left( \frac{\partial u_j}{\partial x_j} \right)^2 \right]
\nonumber\\
&= 2\mu \frac{\partial^2}{\partial x_i \partial x_j}(u_i u_j)
- 4\mu \frac{\partial}{\partial x_i}
\left( u_i \frac{\partial u_k}{\partial x_k} \right)
\nonumber\\
&\quad
+ 2\mu \left( \frac{\partial u_k}{\partial x_k} \right)^2 .
\label{eq:der_product}
\end{align}
Combining eqns. \eqref{eq:viscous_dissi}, \eqref{eq:sijsij}, and \eqref{eq:der_product} yields the decomposition, 
\begin{equation}
\epsilon =  \epsilon_s + \epsilon_d + \epsilon_I\,,
\label{eq:disstotal}
\end{equation}
with the solenoidal (s), dilatational (d), and inhomogeneous (I) components, which follow in vector notation to
\begin{align*}
	\epsilon_s({\bm x},t)  &= \mu ({\bm \nabla} \times {\bm u})^2\,, \;\;
	\epsilon_d({\bm x},t)  = \frac{4}{3} \mu ({\bm \nabla} \cdot {\bm u})^2\,,\\ 
	\epsilon_I({\bm x},t) &=  2\mu \left[({\bm \nabla} \otimes {\bm \nabla}) : ({\bm u} \otimes {\bm u}) - 2\,{\bm \nabla} \cdot \big( {\bm u}({\bm \nabla} \cdot {\bm u}) \big) \right].
\end{align*}
Figure~\ref{fig:isosurface} (a--c) displays isosurfaces of snapshots at $\epsilon = 10\langle\epsilon\rangle$ of the simulations with $M_{\rm rms}=0.1$, $1$ and $10$ and the highest $Re$. We see that $\epsilon({\bm x},t)$ crosses over from local small shear layers (panel~a), which are caused by the self-induced strain due to vortex stretching \cite{Hamlington2008}, to extended, stacked shock-generated sheets (panel~c) for the highly supersonic case. Figure \ref{fig:isosurface} (d--f) displays isosurfaces of $\epsilon_s=10\langle \epsilon_s\rangle$ at the same $M_{\rm rms}$ and $Re$. We see that the solenoidal part $\epsilon_s$ changes from tube-like patterns in panel (d), which are typical for the local enstrophy in the incompressible case \cite{Schumacher2010}, to shock-surface patterns, almost identical to those of the total dissipation $\epsilon$ at $M_{\rm rms}=10$, cf.~panels~(c) and~(f). This qualitative transition of the highest-amplitude dissipation structures from vortex filaments with box counting dimension $D=1$ for $M_{\rm rms} < 1$ to sheet-like structures with $D=2$ at $M_{\rm rms} \gg 1$ is consistent with earlier studies based on velocity structure function scaling~\cite{PJNB2004,SFK2008}.
%---------------------------------------------
\begin{figure}
\centering
\includegraphics{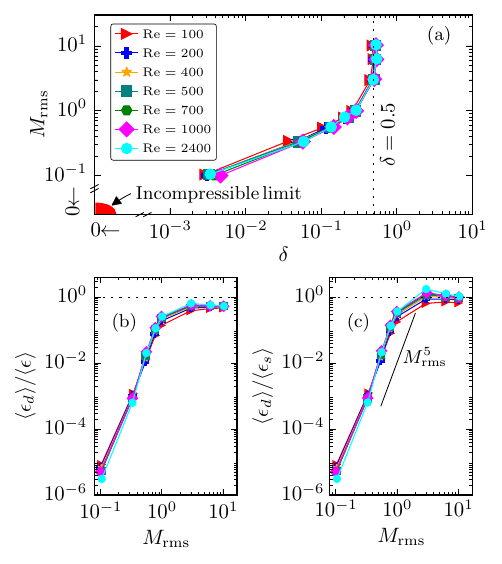}
\caption{(a) Operating points of the solenoidally forced compressible turbulence simulations in the $\delta$--$M_{\rm rms}$ parameter plane (panel~a). The incompressible limit and the $\delta$-asymptote are indicated, which is consistently not exceeded by our simulations. Panels~(b) and~(c) show the ratios of the mean dissipation rates versus the rms Mach number $M_{\rm rms}$: (b) $\langle \epsilon_d \rangle/\langle \epsilon \rangle$ and (c) $\langle \epsilon_d \rangle/\langle \epsilon_s \rangle$. The dotted horizontal lines indicate asymptotic behavior toward a 1:1~ratio.}	
\label{fig:delta_dissi_Mt}
\end{figure}
%---------------------------------------------
\section{Operating points in parameter space} 
We first determine the operating points of our DNS runs in the parameter space of compressible turbulence. In addition to the two central parameters, $Re$ and $M_{\rm rms}$, a third relevant parameter is the dilatational parameter 
\begin{equation}
\delta = \frac{u_{d,{\rm rms}}}{u_{s, {\rm rms}}}\,,
\end{equation}
which quantifies the ratio of dilatational to solenoidal motions in compressible flow \cite{Donzis2020}. Figure~\ref{fig:delta_dissi_Mt}(a) shows our DNS series with solenoidal forcing in the $\delta$--$M_{\rm rms}$ plane. We see an asymptotic behavior for the supersonic DNS to reach approximately $\delta\simeq 0.5$. This finding is consistent with an earlier study \citep{FederrathEtAl2011}, which yielded an asymptotic ratio of $E_\mathrm{sol}/E_\mathrm{tot}=u^2_{s,{\rm rms}}/(u^2_{s,{\rm rms}}+u^2_{d,{\rm rms}})\simeq0.8\Rightarrow\delta\simeq1/2$ for purely solenoidal forcing in the limit $M_{\rm rms}\to\infty$. Furthermore, it is consistent with a renormalization group analysis, which gave a ratio $u^2_{s, {\rm rms}}/u^2_{d, {\rm rms}}=3$, corresponding to $\delta=1/\sqrt{3}\approx 0.58$ \cite{Staroselsky1990}. 

We show the ratios of $\langle\epsilon_d\rangle / \langle\epsilon\rangle$ and  $\langle\epsilon_d\rangle / \langle\epsilon_s\rangle$ versus $M_{\rm rms}$ in Fig.~\ref{fig:delta_dissi_Mt}~(b,c). Panel (b) shows that the dilatational fraction of dissipation is rising from almost zero to $\simeq 0.5$, underlining the increasing role of compressive motions and shock-driven dissipation with increasing Mach number. At the highest Mach number, $M_{\rm rms} = 10$, the dilatational-to-solenoidal mean dissipation ratio (panel~c) is $\lesssim 1$. The inhomogeneous contribution to the total mean dissipation rate, $\langle \epsilon_I \rangle / \langle \epsilon \rangle $, remains relatively small in magnitude as shown in Fig.~\ref{fig:eps_I} in the appendix. We have seen that the isosurfaces of $\epsilon_s$ at this Mach number also form sharp, sheet-like patterns aligned with shocks, likely indicating vorticity amplification in shock–turbulence interactions; this enhanced vorticity increases solenoidal dissipation and may contribute to the observed magnitude of the ratio of about unity for $M_{\rm rms}>1$. In the subsonic regime for $0.3 \le M_{\rm rms} \le 1$, the ratio follows an approximate $M_{\rm rms}^5$ scaling, which is consistent with ref.~\cite{Wang:PRF2017}; a similar but slightly shallower $M_{\rm rms}^{4.1}$ scaling was reported in \cite{JD2016}.

The rate-of-strain tensor can also be Helmholtz--decomposed into two parts, ${\bm S} = {\bm S}_s + {\bm S}_d$ with ${\bm S}_s = [{ \bm \nabla} {\bm u}_s + ({\bm \nabla} {\bm u}_s)^{T}]/2$ and ${\bm S}_d = [{ \bm \nabla} {\bm u}_d + ( {\bm \nabla} {\bm u}_d )^{T}]/2$. Thus, eq. \eqref{eq:eps_total_1} can be alternatively rewritten as
\begin{align}
	\epsilon({\bm x}, t) = 2 \mu ( {\bm S}_s + {\bm S}_d):({\bm S}_s + {\bm S}_d) - \frac{2 \mu}{3} ({\bm \nabla} \cdot {\bm u}_d)^2\,.
	\label{eq:eps_total_2}
\end{align}
We denote $\epsilon_{\rm inc}=2\mu {\bm S}_s:{\bm S}_s$, the pointwise dissipation rate solely arising from the incompressible motion in the flow. Consequently, $\epsilon = \epsilon_{\rm inc} + \epsilon_{\rm com}$ with a compressible dissipation field that collects the remaining terms in \eqref{eq:eps_total_2}. We find that the mean values  $\langle \epsilon_{\rm inc} \rangle$ and $ \langle \epsilon_{\rm com} \rangle$ behave similarly to $\langle \epsilon_s \rangle$ and $\langle \epsilon_d \rangle$ in Fig.~\ref{fig:delta_dissi_Mt}, respectively. Although $\epsilon_{\rm inc}$ and $\epsilon_s$ both arise from solenoidal motions, they are conceptually distinct in their local structure: the former is associated with local shear layers, whereas the latter with the vortical structures in the fluid.

\section{Statistics and scaling of dissipation moments}

\begin{figure}
	\begin{center}
		\includegraphics{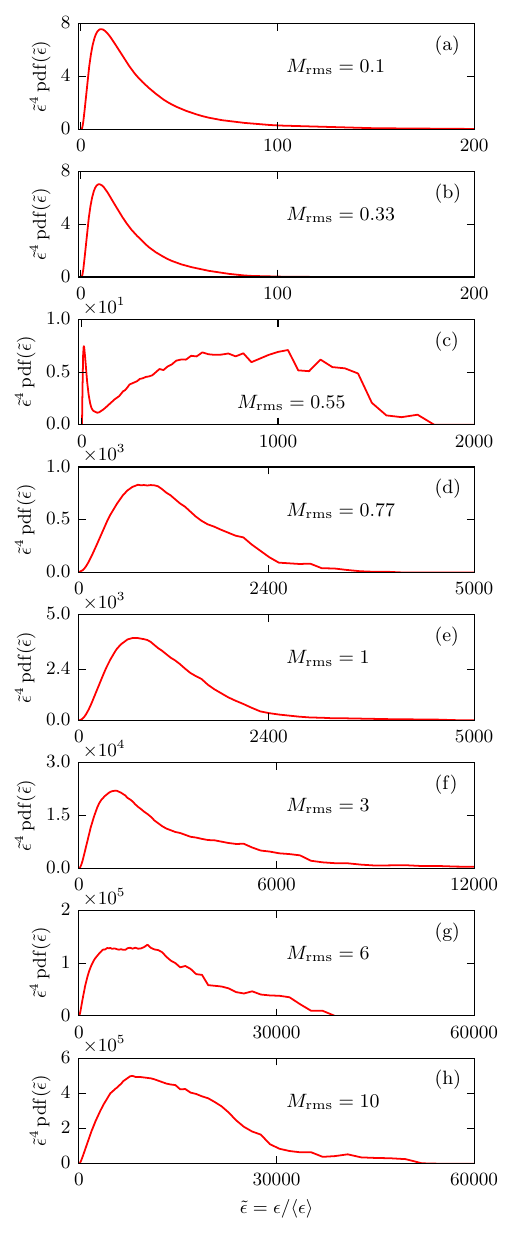}
	\end{center}
	\caption{Statistical convergence test of the normalized 4th-order moments of the total energy dissipation rate field for the highest Reynolds number runs ($Re \approx 2400$) at rms Mach numbers $M_{\rm rms} = 0.1$ (a), $M_{\rm rms} = 0.33$ (b), $M_{\rm rms} = 0.55$ (c), $M_{\rm rms} = 0.77$ (d), $M_{\rm rms} = 1$ (e), $M_{\rm rms} = 3$ (f), $M_{\rm rms} = 6$ (g), $M_{\rm rms} = 10$ (h). }
	\label{fig:convergence}
\end{figure}

\begin{figure}
	\begin{center}
		\includegraphics{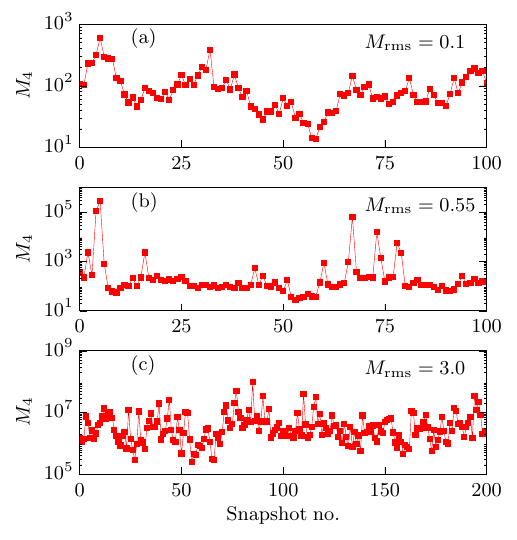}
	\end{center}
	\caption{$4$th-order moments of the total dissipation rate at rms Mach numbers $M_{\rm rms} = 0.1$ (a), $M_{\rm rms} = 0.55$ (b), and $M_{\rm rms} = 3.0$ (c).}
	\label{fig:timeseries}
\end{figure}

Figure~\ref{fig:convergence} illustrates the convergence of the 4th-order moments of the total energy dissipation rate $\epsilon$. We thus plot $\tilde{\epsilon}^4 \, {\rm pdf}(\tilde{\epsilon})$ versus $\tilde{\epsilon} = \epsilon/\langle \epsilon \rangle$ for all Mach numbers $M_{\rm rms}$ at the highest Reynolds number of $Re\approx 2400$. The panels show that all cases are converged, except for the data at $M_{\rm rms}=0.55$. This series is characterized by a few extreme fluctuation events, which alter the overall statistics. Figure~\ref{fig:timeseries} displays the normalized 4th-order moments $M_4(t)$ of total dissipation rate obtained from the different snapshots at $M_{\rm rms}=0.55$ and compares it with those at $M_{\rm rms}=0.1$ and $M_{\rm rms}=3$. First evolving dominant pre-shocks mature to an increasing number of shocks fronts with increasing $M_{\rm rms}$. In Tables~\ref{tab:tab1} and~\ref{tab:tab1a} of the appendix, we provide minima and maxima of $M_4$ for all data series.

%-----------------------------------------------
%-----------------------------------------------
%-----------------------------------------------

%\section{Scaling of dissipation moments} 

%---------------------------------------------
\begin{figure*}
	\begin{center}
	\includegraphics{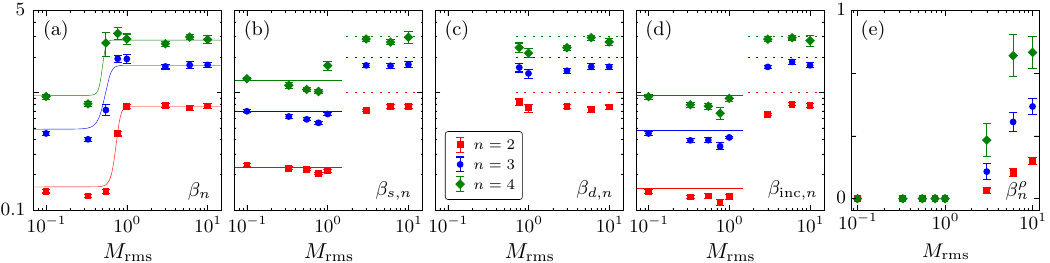}
	\end{center}
	\caption{Reynolds number-scaling exponents of the 2nd-, 3rd- and 4th-order normalized moments versus rms Mach number $M_{\rm rms}$. (a) Total energy dissipation, (b) solenoidal, (c) dilatational, as well as (d) incompressible component of energy dissipation, and (e) density. In the incompressible limit $M_{\rm rms} \ll 1$, $\beta_{{\rm inc}, n} \approx \beta_n$ and are consistent with those reported for incompressible isotropic turbulence~\citep{YS2004,yakhot2006,SSY07}; $\beta_2  = 0.157$, $\beta_3 = 0.489$ and $\beta_4 = 0.944$ . In this limit, the solenoidal exponents (b) take the values $\beta_{s,2} = 0.242$, $\beta_{s,3} = 0.693$, and $\beta_{s,4} = 1.31$, in close agreement with \citet{Elsinga:JFM2023} after necessary adjustment, see Appendix~\ref{appen:esntrophy}. For $M_{\rm rms} > 1$, the exponents of the total dissipation rate field and its components are very close to the Burgers' scaling exponents. In panel~(a), the data points are connected by a tanh fitting curve. The solid and dotted reference lines in (b--d) are for incompressible and Burgers turbulence scaling limits, respectively. We exclude all data at $Re=100$ from the fits due to their closeness to the Gaussian to non-Gaussian transition region of derivative statistics \cite{Donzis2017}.}
	\label{fig:expo}
\end{figure*}
%---------------------------------------------
Figure \ref{fig:expo} summarizes the central result of this work. We show the scaling exponents $\beta_{k, n}$ versus $M_{\rm rms}$ for the time-averaged normalized energy dissipation moments $M_{k,n}(Re)$ of orders $n=2,\, 3$ and $4$. Here, $k=\{s,d,{\rm inc}\}$ in panels~(b--d); panel~(a) shows the total dissipation for completeness. In the low-Mach number regime for $M_{\rm rms}\le 0.33$, the total dissipation exponents $\beta_{n}$ in panel (a) align well with theory~\citep{YS2004,yakhot2006} and DNS~\citep{SSY07,Donzis2017} for incompressible limit. Pre-shocks are not yet present~\cite{TF2025} and moments of density $\langle \rho^n\rangle\sim Re^0$, as seen in Fig.~\ref{fig:expo}(e). In the latter panel, we added the scaling exponents $\beta_n^{\rho}$ of normalized density moments $M^{\rho}_n=\langle \rho^n\rangle/\langle \rho\rangle^n\sim Re^{\beta^{\rho}_n}$ for $n\le 4$. The data points for $\beta_n$ and $\beta_{k,n}$ can be connected by fits or straight lines in Fig.~\ref{fig:expo}(a,b,d), which correspond to the incompressible exponents for $M_{\rm rms}\ll 1$.  

For $M_{\rm rms}=0.55$, the density varies already by $\pm 0.5\langle\rho\rangle$ for the highest $Re$ still leaving $\langle \rho^n\rangle\sim Re^0$ up to $M_{\rm rms}\lesssim 1$, as visible in Fig. \ref{fig:expo}(e). For this $M_{\rm rms}$, we quantify high magnitudes of $-({\bm \nabla}\cdot{\bm u})\tau_{\eta}\sim 10$ that increased by 2~orders of magnitude compared to $M_{\rm rms}=0.33$ (all measured in units of Kolmogorov time $\tau_{\eta}$). This indicates a beginning of the transition in scaling, seen first for $\beta_3$ and $\beta_4$, which are now larger than their incompressible counterparts, caused by locally strongly enhanced dilatational motion.

\begin{figure}
\centering
	\includegraphics[scale=0.9]{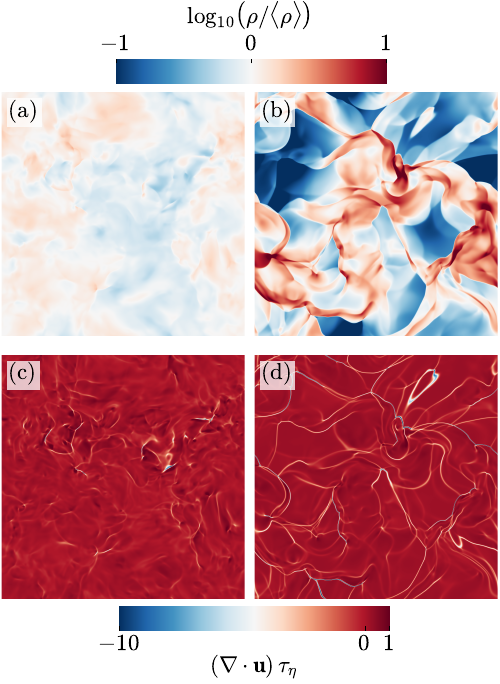}
    \vspace{10pt}
    \caption{Contour slice plots of normalized density and dilatation in the top and bottom row, respectively. (a,c) $M_{\rm rms}=0.77$ and (b,d) $M_{\rm rms}=3$.}
	\label{fig:contours}
\end{figure}

For the subsonic case of $M_{\rm rms} = 0.77$, these dilatation amplitudes $-({\bm \nabla}\cdot{\bm u})\tau_{\eta}$ have been grown to those of the supersonic cases, such that $\beta_n$ and particularly $\beta_{d,n}$ reach a supersonic plateau for the highest orders, see Figs.~\ref{fig:expo}(a,c) and \ref{fig:contours}(a,c). In contrast, the solenoidal and incompressible exponents in Figs.~\ref{fig:expo}(b,d), still remain relatively unchanged up to $M_{\rm rms} \approx 1$, with $\beta_{s,n}$ exceeding $\beta_{{\rm inc},n}$---consistent with findings that $\epsilon_s$ is more intermittent than $\epsilon_{\rm {inc}}$ for $Re_\lambda \lesssim 400$~\citep{Donzis:POF2008,YDS2012}. To conclude this discussion, our DNS suggest that an intermittently emerging dilatational motion for $0.5 \lesssim M_{\rm rms}\lesssim 1$ causes the transition in the small-scale derivative statistics. It is in line with Fig.~\ref{fig:delta_dissi_Mt}(b), which shows $\langle\epsilon_d\rangle/\langle \epsilon\rangle\to {\cal O}(1)$. For $M_{\rm rms}\gtrsim 3$, the density moments $M_n^{\rho}$ transition to an anomalous $Re$-scaling; see Fig.~\ref{fig:expo}(e). Mature shocks in $\rho$ and steepest velocity gradients are then aligned in a shock- or dilatation-dominated turbulence; see Figs.~\ref{fig:contours}(b,d).   

What determines or bounds the plateau value for the $\beta$'s in all panels? In one-dimensional, stochastically forced Burgers turbulence, it was shown by Friedrich et al.~\cite{Friedrich2018} that the normalized moments $\tilde{M}_n$ of the derivative $\partial_x u$ satisfy an anomalous scaling relation for $Re\gg 1$, which is given by and can be translated to, 
\begin{align}
	\tilde{M}_n = \frac{\langle (\partial_x u)^n\rangle}{\langle (\partial_x u)^2\rangle^{n/2}} \sim Re^{n/2-1} \;\;\Rightarrow\;\; M_n\sim Re^{n-1}\,,
    \label{eq:Burgers}
\end{align}
using $\epsilon\sim (\partial_x u)^2$. The plateaus of $\beta_{k,n}$ in Fig.~\ref{fig:expo} fall consistently below 1, 2, and 3 for $M_2$, $M_3$, and $M_4$, respectively. This clearly supports the physical picture that the anomalous exponents of shock-dominated Burgers turbulence (with pressure being absent) bound those of fully compressible turbulence from above.    
  
%---------------------------------------------
%---------------------------------------------

\section{Conclusion} 

Our numerical simulations of compressible turbulence report a transition of the velocity derivative statistics, here in the form of scaling laws for the kinetic energy dissipation and its components with $Re$, from the scaling for the low-Mach number limit ($M_{\rm rms}\ll 1$) to a Burgers turbulence-type scaling for $M_{\rm rms}\gg 1$, dominated by shocks and dilatation sheets for the density and energy dissipation fields. This relatively sharp transition is initiated by increasing dilatational motion for $0.5\lesssim M_{\rm rms}\lesssim 1$. Both components of dissipation, $\epsilon_s$ and $\epsilon_d$, scale similarly for $M_{\rm rms}>1$, since steepest shear and dilation gradients nearly perfectly coincide, which is shown in Figs.~\ref{fig:isosurface}(c,f) and \ref{fig:contours}(b,d). The present analysis is based on an isothermal formulation; therefore flows of molecular fluids that are thermodynamically isolated or subject to weak cooling only cannot be fully represented. Thus, DNS that include the full thermodynamic evolution will be studied in future work. It remains to be seen if this relatively sharp transition stays intact.   

\vspace{0.5cm}
\noindent
\section*{Acknowledgements}
S.A.~and J.S.~are supported by the European Union (ERC, MesoComp, Grant No. 101052786). Views and opinions expressed however are those of the authors only and do not necessarily reflect those of the European Union or the European Research Council. S.A.~was also partly supported by grant no.~SCHU 1410/31-1 of the Deutsche Forschungsgemeinschaft. C.F.~acknowledges funding provided by the Australian Research Council (Discovery Projects~DP230102280 and DP250101526), and the Australia-Germany Joint Research Cooperation Scheme (UA-DAAD). The authors acknowledge the Gauss Centre for Supercomputing (GCS) e.V.~(https://www.gauss-centre.eu) for funding this work by providing computing time on the GCS Supercomputer SuperMUC-NG at the Leibniz Supercomputing Centre (https://www.lrz.de) within the compute projects pn68ni, pn67la, pr32lo, and pr48pi. Authors further acknowledge GCS Large-scale project~10391, the Australian National Computational Infrastructure (grant~ek9) and the Pawsey Supercomputing Centre (project~pawsey0810) in the framework of the National Computational Merit Allocation Scheme and the ANU Merit Allocation Scheme. In addition, the authors gratefully acknowledge the computing time made available to them on the high-performance computer [Otus, under project hpc-prf-kedrict] at the NHR Center Paderborn Center for Parallel Computing (PC2). This center is jointly supported by the Federal Ministry of Research, Technology and Space and the state governments participating in the National High-Performance Computing (NHR) joint funding program.

\section*{Data Availability}
The \texttt{FLASH} source code used to generate the simulation data can be obtained upon reasonable request through the \texttt{FLASH} Center (https://flash.rochester.edu/site/flashcode). The raw simulation data (full three-dimensional fields) are available from the corresponding author upon reasonable request. All post-processed data and Python scripts used to generate the figures (except Fig. 1) are publicly available on Zenodo~\cite{AFS2026_Zenodo}.

\appendix
\section{Numerical method and simulation parameters}
\label{appen:numerical}
Detailed simulation parameters, information about the resolution ($k_{\rm max}\eta$), as well as maxima and minima of the normalized 4th-order moments $M_4$ of the total energy dissipation rate are listed in Tables~\ref{tab:tab1} and~\ref{tab:tab1a}. We define therefore $M_4(t)=\langle \epsilon^4 \rangle_V / \langle \epsilon \rangle_V^4$, where $\langle \cdot \rangle_V$ denotes a volume average. Furthermore, we provide the number of statistically independent snapshots, $N_{s}$. Successive snapshots are separated by a quarter large-scale eddy-turnover time, $T_e/4$ with $T_e=L_f/u_{\rm rms}$. 

\begin{figure}
	\begin{center}
	\includegraphics{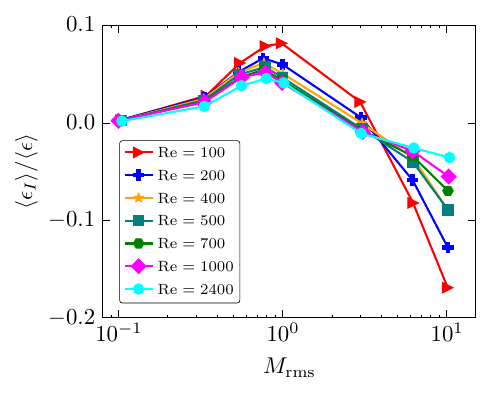}
	\end{center}
	\caption{Inhomogeneous contribution to the total mean dissipation rate as a function of the rms Mach number $M_{\rm rms}$. }
	\label{fig:eps_I}
\end{figure}

We have already discussed the contributions of the solenoidal and dilatational dissipation components to the total dissipation rate at different Mach numbers in Fig.~\ref{fig:delta_dissi_Mt} of the main text. For completeness, we present the inhomogeneous contribution in Fig.~\ref{fig:eps_I}. The figure shows that its magnitude remains relatively small for all Mach numbers. This contribution arises solely from spatial variations of the dynamic viscosity, $\mu=\rho\nu$.
We confirm that the periodicity of the domain yields $\langle \epsilon_I/\mu \rangle \approx 0$. Since $\epsilon_I$ is not positive definite, its contribution may be either positive or negative depending on the local flow structure as seen in the figure.

%--------------------------------------------- 
\begin{table*}
\setlength{\tabcolsep}{8pt}
\centering
%begin{adjustbox}{max width=\textwidth,keepaspectratio}
%\small
\renewcommand{\arraystretch}{0.9}
\begin{tabular}{cccccccccc}
\hline \hline\\[-7pt]
${\rm Run}$ & $N^3$   & $M_{\rm rms}$  & $Re$ & $Re_\lambda$ & $N_s$ & $M_4$ & ${\rm min}[M_4]$ & ${\rm max}[M_4]$ & $k_{\rm max} \eta$ \\[3pt]
\hline \\[-6pt]
\quad $1$  & $512^3$  & $0.10$  & $100$ & $17$ & $100$ & $1.58 \times 10^{1}$ & $1.01 \times 10^{1}$ & $3.77 \times 10^{1}$ & $29.17$ \\
\quad $2$  & $512^3$  & $0.10$  &  $219$& $29$  & $100$ & $2.64 \times 10^{1}$ & $1.51 \times 10^{1}$ & $6.53 \times 10^{1}$ & $17.45$\\
\quad $3$  & $512^3$  & $0.10$ &  $399$& $43$ & $100$ & $4.12 \times 10^{1}$ & $2.29 \times 10^{1}$ & $9.36 \times 10^{1}$ & $11.58$\\
\quad $4$  & $1024^3$  & $0.10$ & $520$ & $50$ & $100$ & $5.33 \times 10^{1}$ & $3.23 \times 10^{1}$ & $1.25 \times 10^{2}$ & $19.24$ \\
\quad $5$  & $1024^3$  & $0.10$ & $709$ & $58$ & $100$ & $6.41 \times 10^{1}$ & $3.90 \times 10^{1}$ & $1.29 \times 10^{2}$ & $15.26$ \\
\quad $6$  & $1024^3$  & $0.10$ & $932$ & $67$ & $100$ & $8.50 \times 10^{1}$ & $4.46 \times 10^{1}$ & $2.63 \times 10^{2}$ & $12.40$ \\	
\quad $7$  & $2048^3$  & $0.10$ & $2469$& $117$ & $100$ & $2.11 \times 10^{2}$ & $9.15 \times 10^{1}$ & $7.69 \times 10^{2}$ & $12.40$ \\[5pt]
\quad $8$  & $512^3$  & $0.34$  & $102$ & $17$  & $100$ & $1.59 \times 10^{1}$ & $8.29 \times 10^{0}$ & $2.94 \times 10^{1}$ & $29.06$  \\
\quad $9$  & $512^3$ & $0.33$ & $222$  & $29$ & $100$  & $2.52 \times 10^{1}$ & $1.63 \times 10^{1}$ & $6.00 \times 10^{1}$  & $17.29$ \\
\quad $10$  & $512^3$ & $0.33$  & $391$ & $42$  & $100$ & $4.51 \times 10^{1}$ & $2.35 \times 10^{1}$ & $1.49 \times 10^{2}$ & $11.74$ \\
\quad $11$  & $1024^3$ & $0.33$ & $511$  & $49$ & $100$  & $5.31 \times 10^{1}$ & $3.21 \times 10^{1}$ & $1.60 \times 10^{2}$ & $19.46$ \\
\quad $12$  & $1024^3$ & $0.33$ & $715$  & $59$ & $100$  & $6.37 \times 10^{1}$ & $3.16 \times 10^{1}$ & $1.59 \times 10^{2}$ & $15.26$ \\
\quad $13$  & $1024^3$  & $0.33$ & $956$  & $70$ & $100$ & $8.85 \times 10^{1}$ & $4.86 \times 10^{1}$ & $6.75 \times 10^{2}$ & $12.39$ \\
\quad $14$  & $2048^3$ & $0.33$ & $2411$ & $116$ & $100$ & $1.69 \times 10^{2}$ & $1.02 \times 10^{2}$ & $3.25 \times 10^{2}$ & $12.64$  \\[5pt]
\quad $15$  & $512^3$  & $0.55$ & $96$ & $16$ & $200$ & $1.80 \times 10^{1}$ & $1.12 \times 10^{1}$ & $3.43 \times 10^{1}$ & $29.88$ \\
\quad $16$  & $512^3$  & $0.54$ & $213$ & $30$ & $200$ & $2.61 \times 10^{1}$ & $1.36 \times 10^{1}$ & $6.58 \times 10^{1}$ & $18.33$ \\
\quad $17$  & $512^3$  & $0.54$ & $378$ & $41$ & $200$ & $5.49 \times 10^{1}$ & $2.29 \times 10^{1}$ & $9.53 \times 10^{2}$ & $12.05$ \\
\quad $18$  & $1024^3$  & $0.55$ & $497$ & $48$  & $200$ & $1.61 \times 10^{2}$ & $2.98 \times 10^{1}$ & $1.81 \times 10^{4}$ & $19.75$ \\
\quad $19$  & $1024^3$  & $0.56$ & $712$ & $61$ & $100$ & $1.89 \times 10^{3}$ & $3.95 \times 10^{1}$ & $1.76 \times 10^{5}$ & $15.51$ \\
\quad $20$  & $1024^3$  & $0.56$  & $955$ & $71$ & $100$ & $6.11 \times 10^{3}$ & $6.57 \times 10^{1}$ & $3.90 \times 10^{5}$ & $12.46$ \\
\quad $21$  & $2048^3$  & $0.56$ & $2405$ & $115$ & $100$ & $5.74 \times 10^{3}$ & $9.87 \times 10^{1}$ & $1.57 \times 10^{5}$ & $12.60$  \\[5pt]
\quad $22$  & $512^3$  & $0.78$  & $97$ & $16$ & $200$ & $5.36 \times 10^{1}$ & $1.11 \times 10^{1}$ & $2.78 \times 10^{3}$ & $29.55$ \\
\quad $23$ & $512^3$  & $0.77$  & $212$ & $29$ & $200$ & $2.91 \times 10^{2}$ & $2.28 \times 10^{1}$ & $1.34 \times 10^{4}$ & $17.91$  \\
\quad $24$  & $512^3$  & $0.77$  & $385$ & $42$ & $200$ & $5.50 \times 10^{3}$ & $4.24 \times 10^{1}$ & $2.09 \times 10^{5}$ & $11.91$ \\
\quad $25$ & $1024^3$  & $0.78$  & $505$ & $49$ & $200$ &  $3.14 \times 10^{4}$ & $9.99 \times 10^{1}$ & $1.11 \times 10^{6}$ & $19.68$ \\
\quad $26$ & $1024^3$  & $0.77$  & $695$ & $59$ & $100$ & $4.58 \times 10^{4}$ & $4.63 \times 10^{2}$ & $6.12 \times 10^{5}$ & $15.68$ \\
\quad $27$  & $1024^3$  & $0.78$ & $947$  & $70$ & $100$ & $7.28 \times 10^{4}$ & $2.91 \times 10^{3}$ & $4.88 \times 10^{5}$ & $12.51$ \\
\quad $28$  & $2048^3$  & $0.78$ & $2414$ & $116$ & $100$ & $1.05 \times 10^{6}$ & $2.64 \times 10^{4}$ & $9.76 \times 10^{6}$ & $12.62$  \\[5pt]
\quad $29$  & $512^3$  & $1.0$  & $95$ & $16$ & $200$ & $1.77 \times 10^{2}$ & $1.56 \times 10^{1}$ & $2.44 \times 10^{3}$ & $29.60$ \\
\quad $30$  & $512^3$  & $1.0$  & $218$ & $29$ & $200$ & $4.10 \times 10^{3}$ & $8.09 \times 10^{1}$ & $5.23 \times 10^{4}$ & $17.40$  \\
\quad $31$  & $512^3$  & $1.0$  & $386$ & $42$ & $200$ & $3.60 \times 10^{4}$ & $5.66 \times 10^{2}$ & $8.06 \times 10^{5}$ & $11.88$ \\
\quad $32$  & $1024^3$  & $1.0$  & $499$ & $49$ & $200$ & $1.35 \times 10^{5}$ & $2.91 \times 10^{3}$ & $1.51 \times 10^{6}$ & $19.78$ \\
\quad $33$  & $1024^3$  & $1.0$  & $686$ & $58$ & $100$ & $3.39 \times 10^{5}$ & $1.03 \times 10^{4}$ & $4.07 \times 10^{6}$ & $15.73$ \\
\quad $34$  & $1024^3$  & $1.0$ & $927$  & $69$ & $100$ & $4.95 \times 10^{5}$ & $2.14 \times 10^{4}$ & $5.72 \times 10^{6}$ & $12.64$ \\
\quad $35$  & $2048^3$  & $1.0$ & $2389$ & $115$ & $100$ &  $4.95 \times 10^{6}$ & $1.10 \times 10^{6}$ & $2.08 \times 10^{7}$ & $12.69$  \\
\hline\hline
\end{tabular}
%\end{adjustbox}
\vspace{0.5cm}
\caption{Direct numerical simulation (DNS) parameters  of solenoidally forced compressible isotropic turbulence. The Run number, rms Mach number $M_{\rm rms}$, large-scale Reynolds number $Re$, Taylor microscale Reynolds number $Re_{\lambda}$, statistically independent snapshots $N_s$, minima and maxima of $M_4$, and $k_{\rm max}\eta$.}
\label{tab:tab1}
\end{table*}
%-----------------------------------------------

%--------------------------------------------- 
\begin{table*}
\setlength{\tabcolsep}{8pt}
\centering
%\begin{adjustbox}{max width=\textwidth,keepaspectratio}
%\small
\renewcommand{\arraystretch}{0.9}
\begin{tabular}{cccccccccc}
\hline \hline\\[-7pt]
${\rm Run}$ & $N^3$   & $M_{\rm rms}$  & $Re$ & $Re_\lambda$ & $N_s$ & $M_4$ & ${\rm min}[M_4]$ & ${\rm max}[M_4]$ & $k_{\rm max} \eta$ \\[3pt]
\hline \\[-6pt]
\quad $36$  & $512^3$  & $3.0$  & $95$ & $15$ & $200$ & $1.70 \times 10^{4}$ & $6.05 \times 10^{2}$ & $3.45 \times 10^{5}$ & $28.32$ \\
\quad $37$  & $512^3$  & $3.0$  & $217$ & $25$ & $200$ & $1.11 \times 10^{5}$ & $9.77 \times 10^{3}$ & $2.25 \times 10^{6}$ & $16.23$  \\
\quad $38$  & $512^3$  & $3.1$  & $390$ & $35$ & $200$ & $4.40 \times 10^{5}$ & $2.93 \times 10^{4}$ & $4.64 \times 10^{6}$ & $10.76$ \\
\quad $39$  & $1024^3$  & $3.0$  & $507$ & $40$ & $200$ & $1.28 \times 10^{6}$ & $9.30 \times 10^{4}$ & $1.47 \times 10^{7}$ & $17.72$ \\
\quad $40$  & $1024^3$  & $3.0$  & $721$ & $49$  & $200$ & $3.27 \times 10^{6}$ & $1.85 \times 10^{5}$ & $8.42 \times 10^{7}$ & $13.92$ \\
\quad $41$  & $1024^3$  & $3.1$ & $944$  & $59$ & $200$ &  $4.44 \times 10^{6}$ & $5.51 \times 10^{5}$ & $6.69 \times 10^{7}$ & $11.31$ \\
\quad $42$  & $2048^3$  & $3.0$ & $2360$ & $96$ & $132$ & $6.01 \times 10^{7}$ & $9.46 \times 10^{6}$ & $5.75 \times 10^{8}$ & $11.74$  \\[5pt]
\quad $43$  & $512^3$  & $6.2$  & $104$ & $16$ & $200$ &$5.45 \times 10^{5}$ & $9.00 \times 10^{3}$ & $3.89 \times 10^{7}$ & $26.94$ \\
\quad $44$  & $512^3$  & $6.2$  & $207$ & $24$ & $200$ & $1.72 \times 10^{6}$ & $2.30 \times 10^{4}$ & $4.47 \times 10^{7}$ & $16.77$  \\
\quad $45$  & $512^3$  & $6.2$  & $416$ & $36$ & $200$ & $1.35 \times 10^{7}$ & $1.98 \times 10^{5}$ & $6.04 \times 10^{8}$ & $10.22$ \\
\quad $46$  & $1024^3$  & $6.2$  & $519$ & $41$ & $200$ & $3.27 \times 10^{7}$ & $6.76 \times 10^{5}$ & $4.77 \times 10^{8}$ & $17.35$ \\
\quad $47$  & $1024^3$  & $6.2$  & $724$ & $49$  & $200$ & $8.33 \times 10^{7}$ & $1.49 \times 10^{6}$ & $2.65 \times 10^{9}$ & $13.64$ \\
\quad $48$  & $1024^3$  & $6.2$ & $1034$  & $59$ & $200$ & $1.54 \times 10^{8}$ & $2.73 \times 10^{6}$ & $8.70 \times 10^{9}$ & $10.49$ \\
\quad $49$  & $2048^3$  & $6.3$ & $2631$ & $97$ & $150$ & $4.88 \times 10^{9}$ & $4.17 \times 10^{7}$ & $4.26 \times 10^{11}$ & $10.49$  \\[5pt]
\quad $50$  & $512^3$  & $10.2$  & $98$ & $15$ & $200$ & $1.50 \times 10^{6}$ & $2.05 \times 10^{4}$ & $8.71 \times 10^{7}$ & $28.42$ \\
\quad $51$  & $512^3$ & $10.2$  & $222$ & $25$ & $200$ & $1.55 \times 10^{7}$ & $2.21 \times 10^{5}$ & $3.62 \times 10^{8}$ & $16.01$  \\
\quad $52$  & $512^3$  & $10.3$  & $396$ & $36$ & $200$ & $5.90 \times 10^{7}$ & $1.29 \times 10^{6}$ & $1.24 \times 10^{9}$ & $10.64$ \\
\quad $53$  & $1024^3$  & $10.2$  & $510$ & $40$ & $200$ & $2.66 \times 10^{8}$ & $3.45 \times 10^{6}$ & $4.56 \times 10^{9}$ & $17.56$ \\
\quad $54$  & $1024^3$  & $10.2$  & $714$ & $48$ & $200$ & $6.93 \times 10^{8}$ & $5.16 \times 10^{6}$ & $2.35 \times 10^{10}$ & $13.69$ \\
\quad $55$  & $1024^3$  & $10.3$ & $958$  & $56$ & $200$ & $1.82 \times 10^{9}$ & $1.40 \times 10^{7}$ & $1.69 \times 10^{11}$ & $11.08$ \\
\quad $56$  & $2048^3$  & $10.4$ & $2445$ & $92$ & $146$ & $1.10 \times 10^{10}$ & $2.83 \times 10^{8}$ & $1.93 \times 10^{11}$ & $11.12$  \\
\hline\hline
\end{tabular}
%\end{adjustbox}
\vspace{0.5cm}
\caption{Continued from Table I.}
\label{tab:tab1a}
\end{table*}
%-----------------------------------------------

%-----------------------------------------------
%-----------------------------------------------

We perform the direct numerical simulations using a modified version of the \texttt{FLASH} code~\citep{Fryxell:2000,Dubey:2008}, based on release~4.0.1, which employs the MUSCL–Hancock HLL5R scheme~\citep{Bouchut:2010,Waagan:2011}. 
\begin{itemize}
\item The MUSCL–Hancock reconstruction provides second-order accuracy in space and time. Cell-averaged primitive variables (density, pressure, and velocity components) are reconstructed with piecewise-linear slopes computed from neighboring cells and limited with the MC (monotonized central) slope limiter. The limiter suppresses spurious oscillations near discontinuities while retaining second-order accuracy in smooth regions.

\item The resulting states are used by the HLL5R Riemann solver to compute the fluxes at the cell interfaces. The scheme is applied in a directionally split manner, where each timestep consists of successive sweeps along the coordinate directions with the fluxes computed using a one-dimensional Riemann solver. This solver belongs to the relaxation-based HLL family of approximate Riemann solvers, ensuring positivity of density and pressure~\citep{Bouchut:2010,Waagan:2011}. The viscous terms in the Navier–Stokes equations are discretized using central finite differences.

\item Time integration is explicit, the timestep is determined by the Courant–Friedrichs–Lewy (CFL) and diffusive stability conditions, whichever is more restrictive. The present version of \texttt{FLASH}~\citep{Cielo:2021,Federrath:2021} is optimized for large-scale simulations by applying the hybrid-precision scheme \citep{FederrathOffner2025}: fluid variables are stored in single precision, while critical operations are performed in double precision. This approach preserves the accuracy of double-precision computations, while reducing computational cost, communication overhead, and memory usage~\cite{Federrath:2021,FederrathOffner2025}.

\item Like all Godunov-type upwind finite-volume methods, the scheme introduces numerical dissipation near discontinuities. However, all simulations are highly resolved and satisfy $k_{\rm max}\eta > 10$. Thus numerical dissipation remains minimal.

\item Another known limitation concerns the asymptotic low-Mach-number limit ($M \rightarrow 0$), where Godunov-type upwind schemes yield incorrect scaling of density and pressure fluctuations. In this limit, the continuous governing equations yield pressure fluctuations that scale as $O(M^2)$, whereas standard Godunov-type discretizations produce larger fluctuations of order $O(M)$. This issue has been discussed extensively (see refs.~\cite{GV1999,GM2004,D2010,CYX2018,CLLYG2022,LBAHEWKR2022,WFBK2025} and references therein), where several techniques, such as preconditioning or modified dissipation, have been proposed and shown to recover the correct $O(M^2)$ scaling. These studies further show that the discrepancy in pressure fluctuations between standard Godunov schemes and their corrected variants becomes negligible at $M = 0.1$. In particular, ref.~\cite{CLLYG2022} examined several Godunov-type schemes, including Roe, HLL, HLLC, Rusanov, and AUSMPW+, and found that the pressure fluctuations predicted by these schemes become nearly indistinguishable at $M = 0.1$ and very close to those obtained with low-Mach-corrected formulations (see Fig.~3 of ref.~\cite{CLLYG2022}). Increasing spatial accuracy further alleviates this issue; for example, ref.~\cite{CYX2018} shows that second-order MUSCL reconstruction allows standard Godunov-type schemes to capture correct pressure fluctuations down to $M = 10^{-2}$ (see Fig.~7 of ref.~\cite{CYX2018}). Since the present simulations also employ MUSCL reconstruction, the lowest Mach number case considered here ($M_{\mathrm{rms}} = 0.1$) lies outside the regime where the low-Mach pressure scaling problem becomes significant.

\end{itemize}
Still, we performed several checks to verify that the present results are not affected by numerical artifacts in the subsonic regime and also in the supersonic regime. The validation results are discussed in the next section.

\section{Validation of turbulence statistics with other benchmark DNS}
\label{appen:validation}
\begin{figure*}
	\begin{center}
	\includegraphics{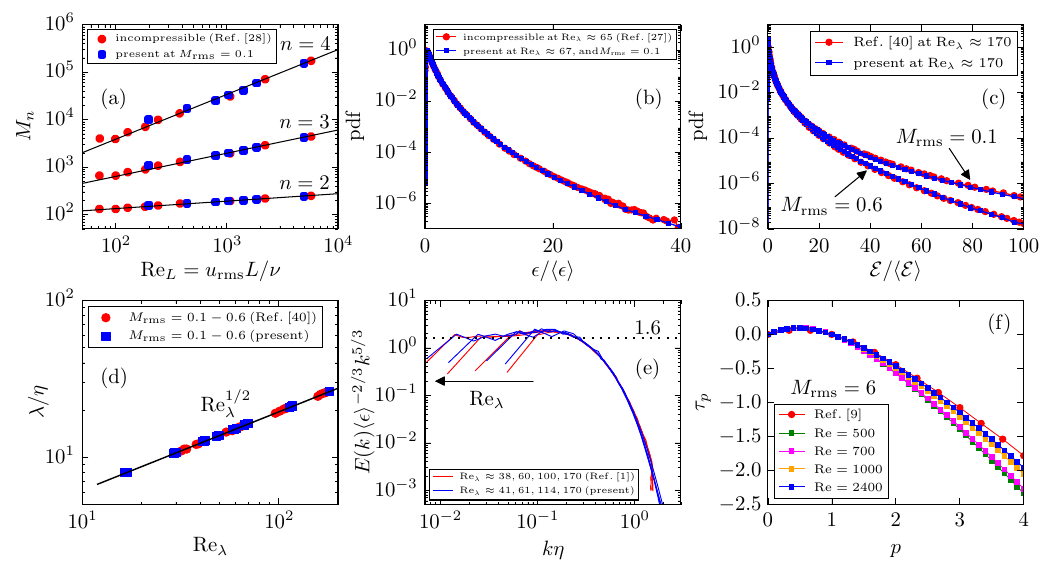}
	\end{center}
	\caption{Validation of the present simulations. Panel (a) compares the Reynolds number dependence of moments of the total energy dissipation rate $M_n$ for orders $n= 2\,, 3$ and $4$ at $M_{\rm rms}=0.1$ with incompressible isotropic turbulence from ref.~\cite{SSK+2014}. Both datasets are vertically shifted by constant factors. Panel (b) shows the pdf of the normalized dissipation rate $\epsilon/\langle\epsilon\rangle$ at $M_{\rm rms}=0.1$ compared with ref.~\cite{SSY07}. Panel (c) shows the pdf of normalized enstrophy $\mathcal{E}/\langle\mathcal{E}\rangle$ at $M_{\rm rms}=0.1$ and $0.6$ compared with ref.~\cite{JD2016}. Panel (d) compares the dependence of $\lambda/\eta$ on $Re_\lambda$ with ref.~\cite{JD2016}. Panel (e) shows the normalized energy spectra at several $Re_\lambda$, compared with ref.~\cite{DJ2013}. Panel (f) compares the scaling exponents $\tau_p$ of the coarse-grained dissipation moments $\langle \epsilon_l^p \rangle \sim l^{\tau_p}$ at $M_{\rm rms}=6$ with ref.~\cite{PPK2009}.}
	\label{fig:validation}
\end{figure*}

The results of the present simulations are not significantly affected by numerical artifacts, as verified by comparison with well-established results from previous studies in Fig.~\ref{fig:validation}. Panel (a) compares the scaling of the moments of the total energy dissipation rate $M_n=\langle \epsilon^n \rangle / \langle \epsilon \rangle^n$ at $M_{\mathrm{rms}}=0.1$ with incompressible homogeneous isotropic turbulence obtained using pseudospectral DNS~\cite{SSK+2014}. The scaling of the present results agrees very well with the incompressible reference data. Panel (b) shows the probability density function (pdf) of the normalized dissipation rate $\epsilon/\langle\epsilon\rangle$ at $M_{\mathrm{rms}}=0.1$ and $Re_\lambda \approx 65$, compared with incompressible DNS results~\cite{SSY07} at the same $Re_\lambda$, again showing excellent agreement.

Panels (c)–(e) provide further validation in the subsonic regime $0.1 \le M_{\mathrm{rms}} \le 0.6$ through comparison with compressible turbulence simulations of refs.~\cite{DJ2013,JD2016}, which employed a tenth-order compact finite-difference scheme. Panel (c) shows pdf of the normalized enstrophy $\mathcal{E}/\langle\mathcal{E}\rangle$ for $M_{\mathrm{rms}}=0.1$ and $0.6$ at $Re_\lambda \approx 170$. For this validation, we carried out two additional simulations at these Mach numbers with $Re_\lambda \approx 170$. The present results agree very well with the reference data. Panel (d) compares the ratio of the Taylor microscale to the Kolmogorov length scale $\lambda/\eta$ as a function of the Taylor microscale Reynolds number $Re_\lambda$. The results follow the expected scaling $\lambda/\eta \sim Re_\lambda^{1/2}$ and agree with the data reported in ref.~\cite{JD2016}. The normalized energy spectra at several $Re_\lambda$ are displayed in panel (e). The spectra agree closely with the compact-scheme results of ref.~\cite{JD2016}. The dependence of the spectra on Mach number in the range $0.1 \le M_{\mathrm{rms}} \le 0.6$ is negligible, apart from minor differences near the forcing wavenumber (not shown).

Panel (f) compares the scaling exponents $\tau_p$ of the moments of the coarse-grained total dissipation rate $\langle \epsilon_l^p \rangle \sim l^{\tau_p}$ at $M_{\rm rms}=6$ with results obtained from simulations of the Euler hydrodynamic equations reported in ref.~\cite{PPK2009} at the same Mach number. The analysis is carried out in the same manner as in ref.~\cite{PPK2009}. The scaling exponents obtained in the present simulations approach the reference data as the Reynolds number increases and are already very close to the reference values at $Re \approx 2400$. In ref.~\cite{PPK2009}, the Reynolds number is controlled by numerical dissipation and is therefore not known explicitly.
%-----------------------------------------------
%-----------------------------------------------

Furthermore, we analyze the pressure spectrum $E_p(k)$, which is theoretically predicted to exhibit a $k^{-7/3}$ scaling in the inertial range of weakly compressible flows~\cite{Bayly1992}. The spectrum is defined as
\begin{align}
E_p(k)\, dk = \sum_{k \le |\bm{k}'| < k+dk}
\left|\hat{p}(\bm{k}')\right|^2,
\end{align}
where $\bm{k}'$ is the wavenumber vector and $\hat{p}(\bm{k}')$ is the Fourier coefficient of the pressure field. At a Reynolds number of $Re_\lambda \approx 350$, ref. \cite{Wang:PRF2017} reported a $k^{-7/3}$ scaling range for $0.015 \lesssim k\eta \lesssim 0.04$ at low Mach numbers ($0.1 \le M_{\rm rms} \le 0.3$). For higher Mach numbers of $0.5 \le M_{\rm rms} \le 1$, the pressure spectrum approached a $k^{-5/3}$ scaling. Also at lower Taylor microscale Reynolds numbers $Re_\lambda \lesssim 100$, ref. \cite{DJ2013} observed a narrow plateau in the compensated spectrum $k^{7/3}E_p(k)$ for the wavenumber range $0.1 \lesssim k\eta \lesssim 0.3$ at rms Mach numbers $M_{\rm rms}=0.1$ and $0.6$.

\begin{figure}
	\begin{center}
	\includegraphics{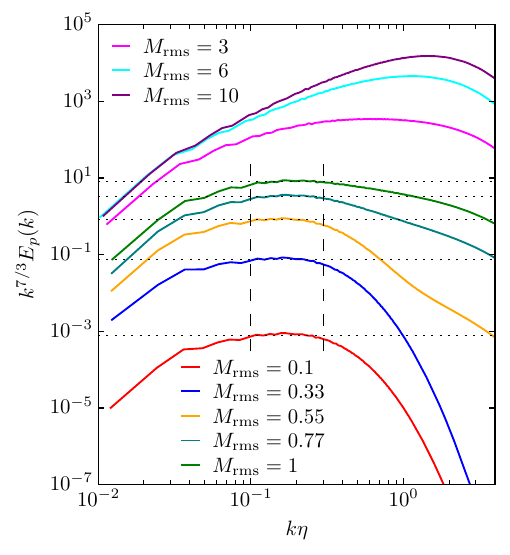}
	\end{center}
	\caption{Compensated pressure spectrum $k^{7/3} E_p(k)$ versus the normalized wavenumber $k\eta$ for simulations of $2048^3$ grid points ($Re_{\lambda} \approx 100$). For $M_{\rm rms}\le 1$, the compensated spectra exhibit a plateau over the interval $0.1 \lesssim k\eta \lesssim 0.3$, indicated by the vertical dashed lines. The horizontal dotted lines show the corresponding plateau amplitudes. For $M_{\rm rms}>1$, the pressure spectrum is shallower than $k^{-7/3}$.}
	\label{fig:pressure_sprectrum}
\end{figure}

Figure~\ref{fig:pressure_sprectrum} shows our own results. We plot the compensated pressure spectrum $k^{7/3} E_p(k)$ for our highest Reynolds number simulations at $Re_\lambda \approx 100$. For $M_{\rm rms}\leq 1$, the compensated spectra likewise exhibit a plateau over $0.1 \lesssim k\eta \lesssim 0.3$, consistent with the findings of \cite{DJ2013}. In comparison with the $k^{-7/3}$ scaling range reported in \cite{Wang:PRF2017}, the plateau appears at larger values of $k\eta$, possibly due to the lower Reynolds numbers considered here. Simulations at higher Reynolds numbers will determine the extent of the predicted $k^{-7/3}$ scaling range. We leave this as a future task. For $M_{\rm rms}>1$, the compensated spectra become progressively shallower than $k^{-7/3}$; this scaling no longer exists as seen in the figure.

%-----------------------------------------------
%-----------------------------------------------

\section{Scaling exponents of enstrophy moments}
\label{appen:esntrophy}
For incompressible turbulence, \citet{Elsinga:JFM2023} investigated the anomalous scaling of enstrophy moments
with respect to the Taylor microscale Reynolds number $Re_{\lambda}$. In their analysis, the scaling was characterized using the $n$th-order moment raised to the power $1/n$, defined as
\begin{align}
    \tilde{M}_{{\rm ens}, n} = \frac{\langle \mathcal{E}^n \rangle^{1/n}}{\langle \mathcal{E} \rangle}  \sim Re_{\lambda}^{\tilde{\beta}_{{\rm ens}, n}}.
\end{align}
This definition can be expressed with respect to the large-scale Reynolds number $Re$ and written directly in terms of the normalized $n$th-order moment, without taking the $1/n$, i.e.,
\begin{align}
  M_{{\rm ens}, n} = \frac{\langle \mathcal{E}^n \rangle}{\langle \mathcal{E} \rangle^n}  \sim Re^{\beta{{\rm ens}, n}} \quad {\rm where} \quad \beta_{{\rm ens},n} =   \frac{n\,\tilde{\beta}_{{\rm ens}, n}}{C}\,. 
\end{align}
Here, we have used the relation \(Re \sim Re_{\lambda}^{C}\) reported by \citet{Elsinga:JFM2023},
based on a fit to the DNS data of \citet{SSY07} with the fit coefficient $C=1.8$. The resulting scaling exponents are listed in Table~\ref{tab:tab2}. In the incompressible limit ($M_{\rm rms} \ll 1$), the dynamic viscosity $\mu = \rho \nu$ is constant, and the solenoidal dissipation component $\epsilon_s$ is therefore directly proportional to the local enstrophy. Consequently, the scaling exponents of the moments of $\epsilon_s$ coincide with the enstrophy exponents, i.e. $\beta_{s, n} \approx \beta_{{\rm ens},n}$, see again Fig.~\ref{fig:expo}(b) in the main text.
%---------------------------------------------------
\begin{table}
\setlength{\tabcolsep}{5pt}
\centering
\small
\renewcommand{\arraystretch}{0.9}
\begin{tabular}{cccccccccc}
\hline \hline\\[-7pt]
\quad $n$ & $\tilde{\beta}_{{\rm ens}, n}$   & $\beta_{{\rm ens}, n}$  \\[3pt]
\hline \\[-6pt]
\quad $2$  & $0.21$  & $0.23$   \\
\quad $3$  & $0.42$  & $0.70$   \\
\quad $4$  & $0.57$  & $1.27$   \\
  \\
\hline\hline
\end{tabular}
\vspace{0.5cm}
\caption{Scaling exponents of enstrophy moments in incompressible homogeneous isotropic turbulence, as reported in a DNS study~\cite{Elsinga:JFM2023}.}
\label{tab:tab2}
\end{table}

%\noindent
%The authors report no conflict of interest.

\def\apj{{\sl Astrophys.~J.}}
\def\apjl{{\sl Astrophys.~J.}}
\def\apjs{{\sl Astrophys.~J.~Suppl.~Ser.}}
\def\aap{{\sl Astron.~Astrophys.}}
\def\aaps{{\sl Astron.~Astrophys.~Suppl.~Ser.}}
\def\aj{{\sl Astron.~J.}}
\def\araa{{\sl Annu.~Rev.~Astron.~Astrophys.}}
\def\mnras{{\sl Mon.~Not.~R.~Astron.~Soc.}}
\def\physrep{{\sl Phys.~Rep.}}
\def\prl{{\sl Phys.~Rev.~Lett.}}
\def\jfm{{\sl J.~Fluid Mech.}}
\def\rmp{{\sl Rev.~Mod.~Phys.}}
\def\jcp{{\sl J.~Comput.~Phys.}}

\bibliography{all,federrath}% Produces the bibliography via BibTeX.

\end{document}